\documentclass[12pt]{article}
\usepackage{xcolor}
\usepackage{amsmath}
\usepackage{graphicx}
\usepackage{times}
\usepackage{wrapfig}
\usepackage{geometry}
\usepackage[utf8]{inputenc}
\usepackage{enumitem,amssymb}
\usepackage{ragged2e}
\usepackage{cite}
\newlist{thematic}{itemize}{8}
\setlist[thematic]{label=$\square$}
\usepackage{pifont}

\usepackage{natbib}
\usepackage{lscape}
\usepackage{authblk}
\usepackage{hyperref}
\usepackage{indentfirst}

\newcommand{\solphys}{Solar~Phys.}
\newcommand{\apj}{Astrophys.~J.}
\newcommand{\apjl}{Astrophys.~J.~Letters}
\newcommand{\mnras}{Monthly Notices Royal Astronom. Soc.}

\newcommand{\aap}{Astron.~Atrophys.}
\newcommand{\ssr}{Space Sci.~Rev.}

\newcommand{\pre}{Phys. Rev. E}

\begin{document}
\raggedright
\huge
Helio2024 Science White Paper \linebreak

{ngGONG -- Future Ground-based Facilities for Research in Heliophysics and Space Weather Operational Forecast.}\linebreak
\normalsize
  
\textbf{Principal Author:} Alexei A. Pevtsov (National Solar Observatory)\linebreak	 
 
\textbf{Co-authors:}
V. Martinez-Pillet (National Solar Observatory), 
H. Gilbert (High Altitude Observatory), 
A. G. de Wijn (High Altitude Observatory),
M. Roth (Thuringian State Observatory, Germany), 
S. Gosain (National Solar Observatory), 
L. A. Upton (Southwest Research Institute), 
Y. Katsukawa (National Astronomical Observatory of Japan, Japan), 
J. Burkepile (High Altitude Observatory), 
Jie Zhang (George Mason University), 
K. P. Reardon (National Solar Observatory), 
L. Bertello (National Solar Observatory), 
K. Jain (National Solar Observatory), 
S.C. Tripathy (National Solar Observatory),  
KD Leka (NorthWest Research Associates, Nagoya University)
\linebreak
\vskip 4cm
\textbf{SYNOPSIS:}
Long-term synoptic\footnote{By synoptic, we mean large-scale (full disk), long-term (solar cycle and longer) observations.} observations of the Sun are critical for advancing our understanding of Sun as an astrophysical object, understanding the solar irradiance and its role in solar-terrestrial climate, for developing predictive capabilities of solar eruptive phenomena and their impact on our home planet, and heliosphere in general, and as a data provider for the operational space weather forecast. We advocate for the development of a ground-based network of instruments provisionally called ngGONG to maintain 
critical observing capabilities for synoptic research in solar physics and for the operational space weather forecast.
\newpage
\section{Introduction}
The Sun’s magnetic field shapes the solar wind and powers explosive phenomena
that result in space weather conditions impacting all natural and artificial bodies within the heliosphere, including our home planet Earth, as well as the
Moon and Mars – next stops in our space exploration quest. To understand the
complex physics behind solar activity and to develop realistic models for
predicting the space weather events
requires essential  and continuous observations of 
solar activity
taken in different wavelengths, covering large spatial scales (full disk), and over 
an extended period (one magnetic cycle or longer). Furthermore, the Sun, the solar system and the heliosphere are paradigms of stellar, planetary and astrosphere systems. This enables the development of a much broader understanding of our own star, and building a bridge between heliophysics and stellar astrophysics based on integrated observations. In addition, synoptic observations also enable data-science-based investigations, as they accumulate broad datasets for future research to solve issues that may not be identified at the time when data are acquired. Finally, observations of stochastic energetic events are an essential source of data for developing a complete understanding of the impact of solar activity on our technological society and for the operational space weather forecast.

\section{Current State: World-wide Synoptic Networks}

There are several advantages of a ground-based network compared to a space-borne instrument \citep[e.g.,][]{Pevtsov2016}. These advantages include: lower 
deployment and operational costs; upgradability and serviceability of instrumentation; broader international collaborations; better bandwidth, redundancy and latency in data delivery; 
ongoing calibration and better long-term survivability.
Currently, synoptic observations of the Sun in USA are conducted at two major ground based networks.

{\bf \textit{Solar Observing Optical Network (SOON)}} includes three stations
situated in USA (Holloman Airforce Base, New Mexico), Western Australia
(Royal Australian Air Force base), and Italy (San Vito dei Normanni Air Station). Routine observations include identification
of solar active regions, measurement of their areas, and imaging in H$\alpha$ 
spectral line. 
SOON is part of Solar Electro-Optical Network operated by DoD/USAF \cite[SEON,][]{Fitts.Loftin1993}.


{\bf \textit{NSF's Global Oscillations Network Group}} (GONG) 
is a 6-site network with instruments
located in USA (Big Bear Solar Observatory, California and Mauna Loa
Observatory, Hawai'i), Western Australia (Learmonth), India (Udaipur Solar
Observatory), Spain (Observatorio del Teide, Canary Islands), and Chile (Cerro
Tololo Interamerican Observatory). An additional (7th) engineering site is located in Boulder, Colorado. 
GONG observations are used for studies of solar interior using methods of helioseismology, evolution of magnetic fields in solar photosphere, and solar eruptive phenomena. In addition, GONG magnetograms and H$\alpha$ images are used 
for the operational space weather forecast by the NOAA Space Weather Prediction Center (SWPC), the US Air Force 557$^{th}$ Weather Wing, UK Met Office, and Japan's Space Weather Center operated by the National Institute of Information and Communications Technology (NICT). GONG is designated by NOAA as the National critical infrastructure for operational space weather forecast. GONG research and development is funded by NSF and its operations are currently funded by NOAA.


\section{ngGONG as future ground-based synoptic network}

Both SOON and GONG networks are aging, and need urgent replacement in order to ensure access to these valuable data. It is expected that GONG project will reach its end-of-life in about ten years (FY2030 or so).
Most importantly, these facilities were designed more than three decades ago, do not have capabilities 
to support answering the pivotal science questions and operational needs that have developed over that time. 
To replace GONG, we propose to build a next generation Ground-based solar Observing Network (provisionally named ngGONG).
The ngGONG will consist of six geographically-distributed stations located at international sites with longitudes, weather patterns, and technical expertise selected to provide nearly continuous observations of the Sun for many decades. Initial consideration would be given to current GONG sites. A seventh, engineering site will be located in Boulder, Colorado to test operational performance, test updated calibration schemes, and validate instrumental upgrades.




\begin{figure}[th!]
\begin{center}
\includegraphics[scale=0.24,clip]{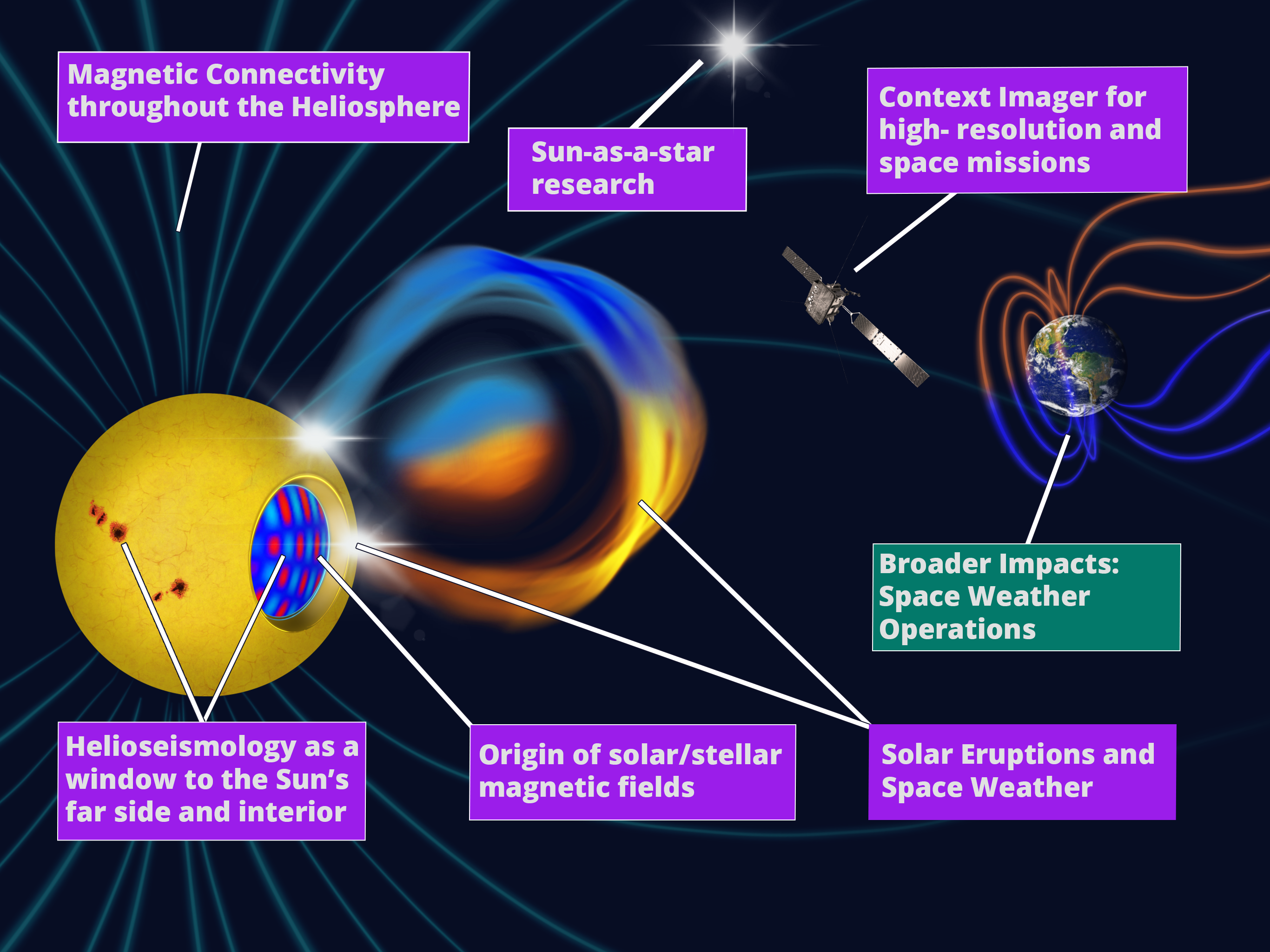}
\caption{Graphical summary of ngGONG science objectives.}
\label{fig:objectives}
\end{center}
\end{figure}

ngGONG will address the following science objectives (SO):

\begin{enumerate}

\item{}
\underline{SO1: Helioseismology as a window into the Sun’s interior and farside.}

The properties of the solar interior constrain models of the solar magnetism that lie at the heart of our interaction with the sun. Large-scale sub-photospheric flows drive the properties of a solar cycle. 
Kinetic helicity of subphotospheric flows has emerged as the critical component of dynamo and even as a proxy of future flare activity of active regions after they crossed the visible solar surface. \citep{Candelaresi.Brandenburg2013,Komm.etal2005}.
As sound waves propagate through the solar atmosphere, they may experience wave transformation into additional magnetohydrodynamic (MHD) oscillation modes (Alfv{\'e}n, fast/slow magnetoacoustic waves, etc.) at heights in the solar atmosphere between the photosphere and chromosphere. Developing a complete understanding of this wave transformation would revolutionize the interpretation of helioseismic signals and open a new window into the properties of the magnetic field in active regions prior to their emergence. 
Helioseismology also offers a “window” to the far-side of the Sun in near real time to forecast the appearance of active regions on the Earth side using acoustic holography \citep{Lindsey.Braun2000} and time-distance techniques \citep{Zhao.etal2019}. These powerful techniques enable an early forecast of medium and large active regions that may have significant influence on space weather throughout the heliosphere. 
Detailed science justifications for helioseismology studies are addressed in  
White Papers by \citet{Jain.etal2022} and \citet{Tripathy.etal2022}.

\item{}
\underline{SO2: Origin of solar/stellar magnetic fields}

The magnetic fields on the Sun and sun-like stars originate in astrophysical dynamos via complex interactions between flows and magnetic fields. 
The evolution of the sub-surface dynamo 
is not well understood and requires long term monitoring of solar magnetic fields and subsurface flows \citep{Elsworth.etal2015}. 
Furthermore, the magnetic field carries information about helicity, which has gained momentum as a key topic of solar and stellar astrophysics.   Several recently discovered trends in helicity are now awaiting an
in-depth understanding of how they develop and evolve in course of the magnetic solar cycle. Some of these trends do not have any theoretical explanation yet. The observational studies of magnetic helicity do not clearly show the theoretically predicted opposite sign of magnetic helicity on either large or small scales for full cycle 24 \citep{Pipin.etal2019} although recent studies do find variations with solar cycle and location of helicity in active regions, \citep{Park_etal_2020,Park_etal_2021}. This may indicate a major shortcoming in our current models of the solar/stellar dynamo. Similar patterns of helicity have been recently observed in cool stars \citep[][]{Lund.etal2020}. This suggests important similarities in solar and stellar dynamos, which solar observations would help resolve. Untangling these puzzling helicity observations requires systematic observations of vector magnetic fields both in the photosphere and the chromosphere, and ngGONG will provide these currently unavailable observations.
For a detailed science justification for vector magnetography with ngGONG see White Paper by \citet{Bertello.etal2022}.
\item{}
\underline{SO3: Science Objective 3: Solar Eruptions and Space Weather.}
 
The Sun’s magnetic field is the primary energy source for all solar eruptive phenomena. The magnetic field also governs the stability of magnetic field configurations, and its changes often serve as a trigger for solar eruptions.
Thus, highly sensitive measurements of the vector magnetic fields in the photosphere and chromosphere are essential to estimate the amount of twist (helicity) and free energy is stored in active region magnetic fields. The information is important for understanding the mechanisms of how the field becomes unstable and to predict the magnitude of a flare \citep{Kusano.etal2020}.
Chromospheric filaments (prominences at the limb) often form the core of a Coronal Mass Ejection (CME). Thus, measuring the vector magnetic field would provide critical information about magnetic topology of filaments, which in turn, would enable early forecast of strength of geomagnetic effects prior to when these chromospheric structures erupt and form CMEs. During eruption, prominences (and filaments) display rotational motions that depend, in part, on the amount of helicity they possess \citep{Fan2011,Torok.etal2010}. Coronal observations are used to estimate the amount of prominence writhe \citep{Torok.etal2018} and to track non-radial trajectories. White light coronal data are needed to measure CME speeds, widths, and line-of-sight densities to determine the severity of space weather impacts. The ngGONG white light coronagraph (based on Mauna Loa K-Cor) can provide near-real-time CME alerts for space weather predictions (see WP by Burkepile et al. (2022).

A combination of full vector field observations in the photosphere and chromosphere combined with coronal observations from a ChroMag-type Polarimeter would enable transformative modeling of evolution of solar corona due to the changing magnetic background as well as the propagation of magnetized CMEs 
from the solar surface to planetary magnetospheres (see also the White Paper by A.G.~de~Wijn et al. (2022).

 Only through long-term sustained observations can we obtain large-sample data for basic understanding of solar behavior as well as for statistical analysis that leads to, e.g., space-weather forecasting ability.  For example, full-disk continuous observations allow a statistical analysis of sunspot regions and flare activity whose results can be used to constrain models of solar flares, but also lead to flare prediction algorithms \citep{LekaBarnesWagner2018}; in the specific cited case, GONG data are the basis for a parallel system that provides quantitative flare prediction when HMI data are unavailable.
\item{}
\underline{SO4: Magnetic Connectivity throughout the Heliosphere.}

    The Sun’s interior, atmosphere, and heliosphere form a single topologically connected system. A comprehensive understanding of how the system evolves and how its different parts interact with each other on different spatial and temporal scales requires synoptic observations. 
 For modeling the solar wind and magnetic connectivity in heliosphere, magnetic fields extrapolation should include coronal holes. However, the magnetic field in coronal holes, especially near solar poles is weak. To measure solar polar magnetic fields requires a telescope with collecting aperture of 0.5 meter or larger. This is true even for magnetograph instruments on Earth orbit, such as SDO/HMI. ngGONG is designed as 0.5 meter telescope, and to increase it sensitivity, it will take vector magnetic field in infrared part of solar spectra. This combination will enable taking critical measurements of magnetic fields in solar polar regions and provide better boundary conditions for modeling of solar wind and interplanetary magnetic field.
A detailed science justification for role of polar magnetic fields in 
establishing 
magnetic connectivity throughout the Heliosphere is discussed in a White Paper by \citet{Petrie.etal2022} 

 Coronal observations are needed to improve solar wind models, currently driven from photospheric boundary conditions where conditions are dramatically different from the corona. ngGONG will provides unique coronal observations using coronal emission lines 
 (based on Mauna Loa CoMP/UCoMP coronagraph) and white light (based on K-Cor). These observations have been used to improve and validate solar wind models \citep[see][]{Jones.etal2020,Corchado-Albelo.etal2021,Cho.etal2020,Mikic.etal2018}. A network of coronagraphs provides more continuous tracking of coronal evolution and further enhances the scientific potential of these data.


\item{}
\underline{SO5: Sun-as-a-star research.}
 
    The Sun provides a critical benchmark for the general study of stellar structure, magnetic activity, and evolution. Solar astronomy has the advantage of having simultaneous observations of disk-resolved and disk-integrated (Sun-as-a-star) spectra. This allows evaluation of the contribution of various solar features on the disk to the disk-integrated signal. This is a vibrant discovery area in solar-stellar research that will help to develop new approaches for interpreting stellar spectra in the search for exoplanets or studying the patterns of activity that may exist on other stars. 
    Regular Sun-as-a-star observations would also allow for the continuation of long-term records of solar activity taken in different spectral bands. Such records are essential for long-term monitoring/study of solar spectral irradiance - one of a few critical parameters for Earth climate modelers.  A combination of ngGONG Sun-as-a-star and disk-resolved (imaging, spectral, and magnetic) observations will provide a unique dataset allowing to bridge solar and stellar astrophysics.
    
For science justification of Sun-as-a-star research see WP by \citet{Criscuoli.etal2022}.
 
\item{}
\underline{SO6: Context Imager for high-resolution and space missions.}


Routine 24/7 observations from ngGONG will allow investigation of coupling between the processes taking place on different spatial scales and provide support as a context imager to high-resolution instruments, such as the NSF's Daniel K. Inouye Solar Telescope \citep[DKIST,][]{Rimmele.etal2020}, and current and future space missions (e.g., Parker Solar Probe, Solar Orbiter, MUSE, and potential L4/L5 or 4$\pi$ missions). Currently, GONG observations are used routinely as context images and as model input in support of several NASA and ESA missions.  The aging 
of GONG puts the future availability of this important context data into question.
\end{enumerate}

For additional science justification of ngGONG, see White Paper by \citet{Pillet.etall2022}.

\section{Recommendation}
The critical science questions about the long-term evolution of solar activity, the fundamental processes taking place in solar interior and the solar atmosphere, and the magnetic connectivity throughout the heliosphere 
can only be answered by
improving our capabilities for synoptic observations of the Sun. 
The long-term goal is to distribute the network nodes around the world such that the Sun can be nearly continuously monitored, and to equip them with a suite of necessary instrumentation. This instrumentation will provide a variety of observables that will serve the science needs of the solar physics community in answering their fundamental research questions and the operational needs of space weather forecasters by providing robust synoptic observations.
The instruments on ngGONG shall include: spectropolarimeters for the precise measurements of solar magnetic fields at multiple heights; broadband imagers and coronagraphs capable of monitoring the violent ejecta of magnetized plasma from the Sun's atmosphere and determining coronal magnetic topologies and plasma properties; instruments for Doppler velocity measurements required for studies of helioseismology. ngGONG instrumentation shall also allow at least preserving the current capabilities required by the operational space weather, and have a design that allows future upgrades to keep the facilities capable of taking new types of observations as demanded by scientific and operational evolution.


\section{Inter-agency and International collaborations}


ngGONG will provide observations both for research and operational space weather communities. To build and operate a multi-use network of such magnitude the inter-agency collaboration is the must. Some of this work has been started by incorporating the requirements from a variety of stakeholders -- the NSF/Science community, DoD/USAF, and NOAA/SWPC.

As a global network, ngGONG can provide a platform for 
extensive international collaboration, which could include multinational partnerships. In fact, some of these collaborations have already been initiated in the framework of SPRING initiative. 
{\bf \textit{SPRING}} is a European effort, with US and Japanese participation, in the international undertaking of upgrading GONG. So far, SPRING has carried out a scientific requirement study focusing on the observational needs for answering the key questions of solar physics \citep{Roth.etal2016}, and translating these  requirements into a technical concept \citep{Gosain.etal2017}.

\section{ngGONG Construction Cost and Schedule Estimates}

\begin{figure}[h!]
\includegraphics[width=1.25\textwidth,angle=90]{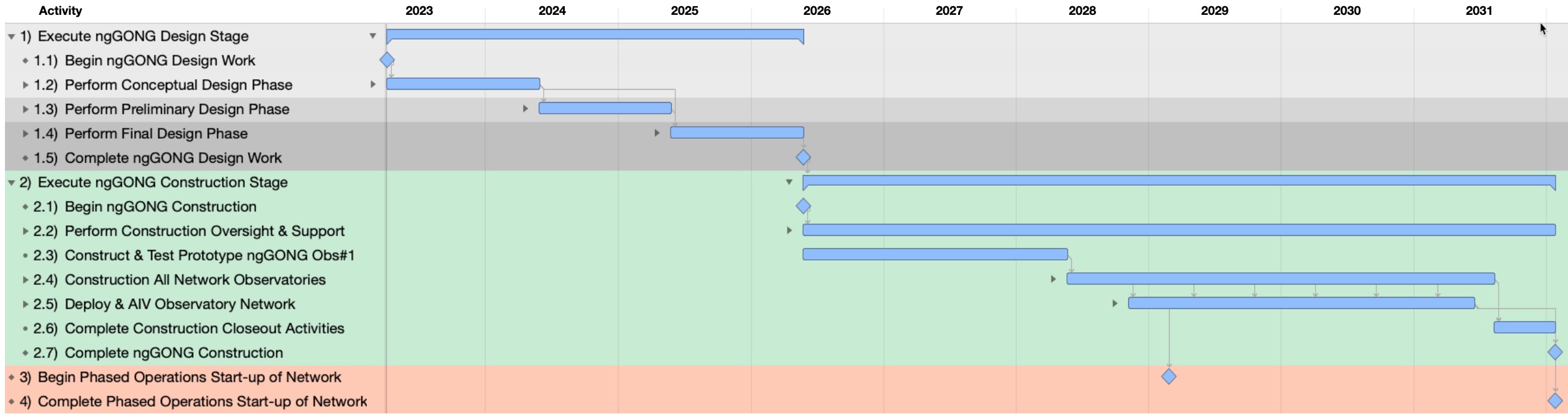}
\includegraphics[width=1.25\textwidth,angle=90]{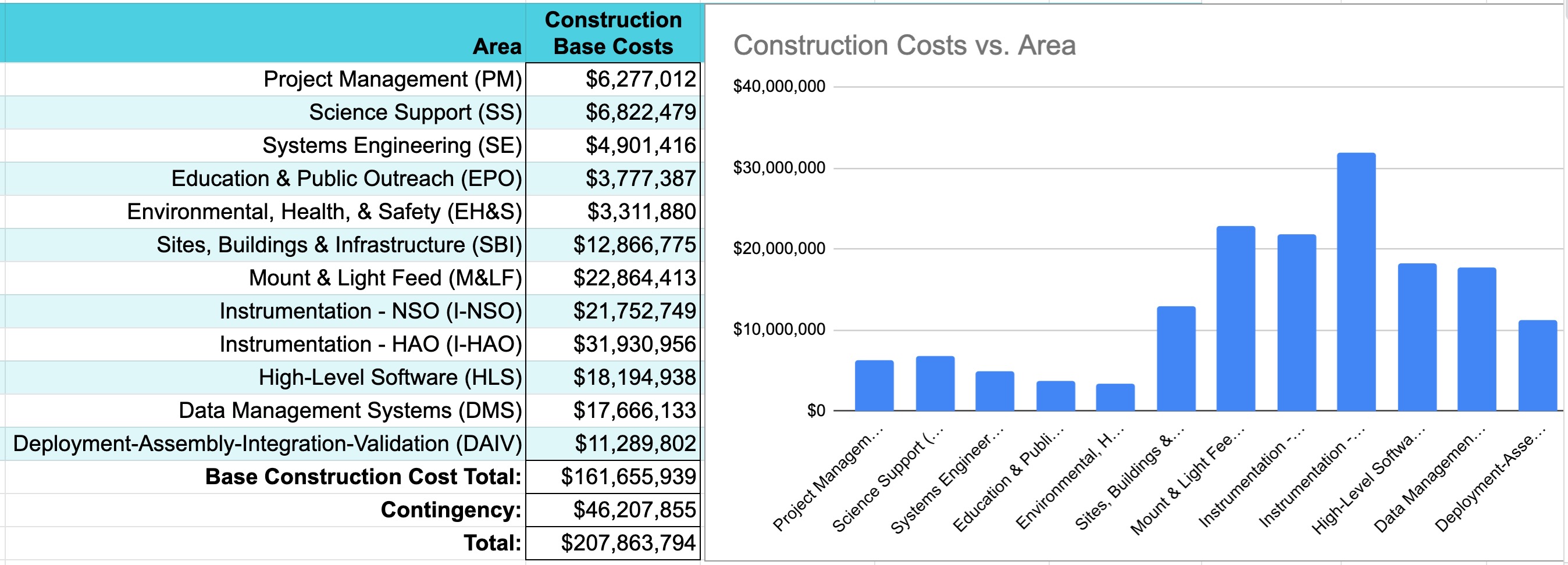}
\caption{(Top): Timeline of 7-site ngGONG network development.
(Bottom): Bottom-up construction cost estimate for 7-site ngGONG network. \label{fig:cost}}
\end{figure}

Construction cost and schedule estimates were made on the basis of a level-3 Work Breakdown Structure (WBS). The ngGONG timeline (Figure \ref{fig:cost}, top) includes approximately three years of design, followed by five and a half years of construction. Both top-down and bottom-up construction cost estimates were generated, yielding similar total construction cost estimates of roughly \$160M for 
seven (7) 
stations (six network stations and one engineering site). The summary of a bottom-up cost estimate by WBS areas is 
shown in Figure \ref{fig:cost} (bottom). Cost estimates for specific areas and instruments were based on labor estimates, and prorated/scaled up/down costs of instruments with similar capabilities built for GEMINI, SOLIS, and DKIST facilities. The estimate assumed that all seven stations will have instruments for helioseismology, broadband imaging and H$\alpha$ Doppler velocity, four out of seven will have full-disk polarimetry instruments in IR (full disk vector magnetic field measurements in the photosphere and chromosphere), three of seven will be equipped with coronagraphic instruments, and four of seven  will have Sun-as-a-star instrumentation. This instrument allocation is driven by science requirements for duty cycle of the different science objectives.
Due to expected uncertainties in our assumptions, a $\sim$30\% contingency was added, thus resulting in a total construction cost of about \$208M. 
Design costs for the combined conceptual, preliminary, and final design phases are currently under development, but are expected to be approximately ten to twenty percent of the total construction costs.
Based on the science/data product requirements, ngGONG is mostly (two thirds or 65\%) a research facility. About one third (35\%) of ngGONG contribution is to provide data for the operational space weather forecast. The cost estimates and assumed contingency values presented here are based on early development stage designs and concepts. We are however confident in the assumed bases of estimates and are comfortable subjecting them to TRACE-type reviews as required.

\newpage


\end{document}